\documentclass[twocolumn,prl,groupedaddress,showpacs,superscriptaddress,amsmath]{revtex4-1}
\usepackage{subfigure}
\usepackage{color}
\usepackage{epsfig,amsfonts,amsbsy,graphicx,amsfonts}

 \def\vec#1{\mbox{\boldmath $#1$}}

\begin{document}

\title{Extracting energy from  non-equilibrium fluctuations  without using information}

 \author{Shou-Wen Wang}
\email{wangsw09@gmail.com}
\affiliation{Beijing Computational Science Research Center, Beijing, 100094, China}
\affiliation{Department of Engineering Physics, Tsinghua University, Beijing, 100086, China}

\author{Lei-Han Tang}
\email{lhtang@csrc.ac.cn}
\affiliation{Beijing Computational Science Research Center, Beijing, 100094, China}
\affiliation{Department of Physics and Institute of Computational and Theoretical Studies, Hong Kong Baptist University, Hong Kong, China}

\date{\today}
  \graphicspath{{./figure/}}
  
\begin{abstract}
Extracting energy from fluctuations has been an everlasting endeavor.  An important strategy is to use information-based operation,  which allows energy extraction even from thermal fluctuations.  However,  it remains  unclear whether a \emph{blind} external  operation, ignorant of such fluctuations, will also work.   Assuming such a  small operation,  we find that  energy extraction requires a negative response spectrum in a certain frequency region,  which  inevitably requires an internal power source to maintain.    Energy is then extracted by operating the system at these frequencies.   Negative response is intrinsic to  any system that  adapts to a slow external operation,   which could be realized through  integral feedback control.   These are demonstrated  through a solvable model. 



\end{abstract}


\maketitle

\emph{Introduction.--} Extracting energy from fluctuations challenges our fundamental understanding of thermodynamics.  Maxwell first showed that  energy could be extracted from thermal fluctuations by an intelligent demon that can detect the velocity of a single particle~\cite{maxwell2012theory}.  The apparent violation of the second law of thermodynamics was clarified later by noting that the energy source is information, and its erasure  is inevitably dissipative~\cite{landauer1961irreversibility,maruyama2009colloquium}.   Recently,  such an information engine has been demonstrated experimentally~\cite{toyabe2010experimental,berut2012experimental,John2014Landauer,Koski2015MaxwellDemon}.  Remarkable theoretical progress has also been achieved along this line~\cite{sagawa2010generalized,mandal2012work,Horowitz2013motor,Horowitz2014information,parrondo2015thermodynamics}.  Another strategy of energy extraction is to exploit   the temperature difference between two baths, by using a single particle.  Such a stochastic heat engine has also been  achieved in recent experiments~\cite{blickle2012realization,Martinez2016Brownian,Rosnagel2016atomengine}.   Its   energetics and efficiency  is under active investigation~\cite{Curzon1975efficiency,Broeck2005efficiency,Verley2014Efficiency,Shiraishi2016efficiency}. Finally,  chemical engines could also be constructed,  as achieved by various molecular motors within the cell~\cite{Schliwa2003Motors}.

Let us  classify these engines based on their operation. 
The stochastic heat engine requires  coupling two different external operations, i.e., the temperature modulation and mechanical control.  So does the chemical engine.   For the case of periodic driving,  their  thermodynamics  have been studied recently~\cite{Brandner2015thermodynamics,Proesmans2015periodic,Proesmans2016periodic,Camerin2016dissipation}.   The information engine, on the other hand,  requires only one external operation,  which  needs to be coupled with  fluctuations  of the system.  It is intriguing whether  a \emph{blind} external  operation,   ignorant of such fluctuations,  will also work.   Although  such  an operation is surely dissipative on passive systems,  as suggested by the second law of thermodynamics,  it remains  unclear for active systems that are driven out of equilibrium by an internal energy source.

Here,  assuming such a small operation,   we find that energy extraction requires a negative response spectrum in a certain frequency region,   and an external operation  at these frequencies.   Negative response is intrinsic to any  system that  adapts to a slow external operation,  which could be  realized through integral feedback control.    These are   demonstrated  in a solvable model.

\emph{General requirements.}--Let us first consider how to  extract  energy from a fluctuating trajectory $x_t$ by applying an external force  $f_t$: 
\begin{equation}
\gamma \dot{x}=F(x,{\vec y},t)+f+\xi.
\label{eq:main-langevin}
\end{equation} 
Here,     $\gamma$ is the friction coefficient,   $\xi$ any stationary noise,  and   $F(x,{\vec y},t)$ an arbitrary  force that could depend on both multiple internal variables   ${\vec y}$ or an internal driving protocol.  The assumption of  an over-damped motion is not important to our main results.   The Boltzmann constant is set to be 1 throughout this paper.  Using $\langle \cdot\rangle_*$  to indicate long-time averaging over $t$,   the energy injection rate through the external operation  is given by 
\begin{equation}
\dot{W}_{ext}^*=\langle   f_t\circ \dot{x}_t\rangle_*\equiv  \lim_{\mathcal{T}\to \infty} \frac{1}{\mathcal{T}}\int_0^\mathcal{T}  f_t\circ \dot{x}_t dt,
\label{eq:Wext}
\end{equation}  
 where $\circ$ denotes  Stratonovich calculus~\cite{sekimoto2010stochastic}. The energy is extracted when  $\dot{W}^*_{ext}<0$,  which is stored in an auxiliary system that connects with $f$.

Let us consider the following  external operation:
\begin{equation}
f_t=(1-\alpha) h_t- \alpha \int_{-\infty}^{\infty } G(t-\tau) \dot{x}_{\tau} d\tau,
\label{eq:ft}
\end{equation}
which mixes a blind perturbation $h_t$,  regardless of $x_t$, and an information-based  operation with explicit dependence on the trajectory, supposed to be generated by  Maxwell's demon.   Since the demon cannot predict the future,   the Green function  satisfies $G(t-\tau)=0$ for $\tau>t$.  Here,  $\alpha\in [0,1]$,  which is a qualitative measure of the  information contained  in the external operation.

To quantify $\dot{W}_*$, we need to introduce the  velocity correlation under the protocol $f_t$: $C_{\dot{x}}^*(\tau)\equiv \langle  (\dot{x}_t-v_0)(\dot{x}_{t+\tau}-v_0)\rangle_*$,   with  $v_0\equiv \langle \dot{x}_t\rangle_*$ the drifting velocity.   For simplicity,  we  take  $v_0=0$,  which  excludes the trivial possibility of  extracting energy with a constant load,  as is the case for  Feynman's ratchet~\cite{Feynman1966lectures}.  The correlation of the perturbation protocol is also important, given by 
$C_h^*(\tau)\equiv \langle (h_t-h^*)(h_{t+\tau} -h^*)\rangle_*$,   assumed to be stationary,   with $h^*\equiv \langle h_t\rangle_*$ the average perturbative force.  
Assuming that $\delta h_t=h_t-h^*$ is a small perturbation to the system,    the  velocity response of the system is characterized  by $  \langle \dot{x}_t\rangle=\int_{-\infty}^\infty R_{\dot{x}}(t-\tau) \delta h_\tau d\tau$.  This   defines a \emph{velocity} response function $R_{\dot{x}}(t-\tau)$,  which is independent of $h_t$,  but relies on the operation of the demon,  which has changed the steady state.  Besides,  $R_{\dot{x}}$   should be  zero for  $t<\tau$ since any perturbation  cannot affect the past response.   Combining the linear response with Eq.~(\ref{eq:Wext}) and Eq.~(\ref{eq:ft}),  we finally obtain  (See Supplemental Material for its derivation and the case with $v_0 \neq 0$~\cite{supp}) 
\begin{equation}
\dot{W}^*_{ext}=\int_{-\infty}^\infty  \Big[ (1-\alpha) \tilde{R}_{\dot{x}}'(\omega) \tilde{C}_h^*(\omega)-\alpha \tilde{G}'(\omega)\tilde{C}_{\dot{x}}^*(\omega)\Big]\frac{d\omega}{2\pi},
\label{eq:spectrum}
\end{equation}
valid up to the second order of $h_t$.  Here, the  tilde  denotes the  Fourier transform and the prime  takes the real part. On the right hand side,    the first part  gives the  frequency-resolved  energy injection rate for a blind perturbation,  which  generalizes a similar result in equilibrium statistical mechanics~\cite{Sethna2006entropy}, while the second part is due to the information-based operation.   It is  reminiscent of the Harada-Sasa equality that connects the steady state dissipation  with the correlation and response spectrum~\cite{harada2005equality}.  Although we have assumed a weak external operation,   Eq.~(\ref{eq:spectrum}) is valid for any systems that are strongly driven by an internal energy supply.  Below,  we discuss    general strategies for energy extraction based on Eq.~(\ref{eq:spectrum}).

Note that   $\tilde{C}_{\dot{x}}^*(\omega)=\omega^2 \langle  |\tilde{x}^+(\omega)|\rangle \ge 0$ and $\tilde{C}_h^*(\omega)=|\tilde{h}^+(\omega)|^2\ge 0$,  with $\tilde{z}^+(\omega)$ a rescaled Fourier transform of a trajectory $z_t$,  given by $\lim_{\mathcal{T}\to\infty} \frac{1}{\sqrt{\mathcal{T}}} \int_0^\mathcal{T} z_t \exp(i\omega t) dt$.   $\tilde{G}'(\omega)$ could also be tuned externally to be non-negative.  Therefore,  when we are implementing only  the information-based operation,  i.e., $\alpha=1$,   we could extract energy even when the system is passive.   This does not violate the second law since more energy will be dissipated  to generate this information-based operation~\cite{maruyama2009colloquium,landauer1961irreversibility}. 

On the other hand,  at $\alpha=0$,  we can extract energy only when the velocity response spectrum $\tilde{R}_{\dot{x}}'(\omega)<0$ at a certain frequency region.   Negative response implies that  the system pushes backwards when we apply a forward external force.  This becomes precise in the frequency domain since   $\langle \tilde{x}'(\omega)\rangle=\tilde{R}_{\dot{x}}'(\omega)\tilde{h}(\omega)$.       Given  such a response spectrum  that is intrinsic to the system and measurable,  we can 
optimize the external perturbation spectrum  accordingly to improve the  output power.

   However,  the equilibrium response is always lagged behind the external operation and energy needs to be dissipated  to ``drag"  the system forward,  as suggested by  the \emph{fluctuation-dissipation theorem} (FDT): $\tilde{R}_{\dot{x}}'(\omega)=\frac{1}{2T}\tilde{C}_{\dot{x}}(\omega)\ge 0$,  with $T$ being the temperature of the bath~\cite{kubo1966fluctuation}.   Here,  $\tilde{C}_{\dot{x}}$ is the correlation spectrum obtained at $h_t=h^*$.  Alternatively,  we may introduce the frequency-resolved effective temperature $
T_{eff}(\omega)=C_{\dot{x}}(\omega)/2R_{\dot{x}}'(\omega)$~\cite{LeticiaPREtemperature}.   Therefore,   a negative effective temperature is required for extracting energy,  which is possible only for far-from-equilibrium systems,  driven by an internal energy source.  

An  insensitive displacement response to low frequency perturbations, i.e.,  $\tilde{R}_{x}'(0)\equiv \tilde{R}_{x}'(\omega)|_{\omega\to 0}\approx 0$,  is sufficient to generate negative response.  To see this,  we need to use the \emph{Kramers-Kr{\" o}nig relation}~\cite{Sethna2006entropy}: $\tilde{R}_x'(\omega)=\frac{2}{\pi} \int_0^\infty \tilde{R}_x''(\omega_1)\frac{\omega_1}{\omega_1^2-\omega^2}d\omega_1$, where the double prime denotes the imaginary part and the integration is to be understood as the principle value.  This relation is simply due to the causality constraint, i.e.,   a perturbation cannot affect responses in the past.  This implies  that  $\tilde{R}_{x}'(0)=\frac{2}{\pi} \int_0^\infty [\tilde{R}_x''(\omega_1)/\omega_1 ]d\omega_1$. To satisfy $\tilde{R}_{x}'(0)\approx 0$,  $\tilde{R}_x''(\omega)$ must have negative response in a certain frequency region to balance  positive response in other frequencies.  So does $\tilde{R}_{\dot{x}}'(\omega)$,  which equals $\omega\tilde{R}_x''(\omega)$.

 Such a response  implies adaptation of the system to a slow external perturbation.  To see this,  let us consider   a sudden jump of the external perturbation $h^*\to h^*+\Delta h$ at $t=0$.   The displacement response satisfies $\langle x_\infty\rangle-\langle x_0\rangle=\Delta h\int_0^\infty R_x(t)dt=\Delta h \tilde{R}_x'(0)\approx 0$, i.e.,  $\langle x_t\rangle$ is recovered after a transient variation.  This implies that   the steady state displacement is invariant under the change of a constant  external force,  called  \emph{adaptation}.

 Adaptation is widely exploited by  sensory systems to recover sensitivity after a transient response to a shift  of external signal~\cite{tu2013quantitative}.  Therefore,   negative response should  be generic in such systems.  This has been revealed in a model for the sensory circuit in E. \emph{coli}
 ~\cite{Pablo2015Adaptation,Wang2016FRRviolation},   which   serves to detect  variation of the extracellular ligand concentration~\cite{tu2013quantitative}.  For the hair bundle  from the  inner ear that senses the auditory  stimuli~\cite{Hudspeth2008hair},  negative response has been revealed experimentally~\cite{martin2001comparison}.

\begin{figure}
\includegraphics[width=8cm]{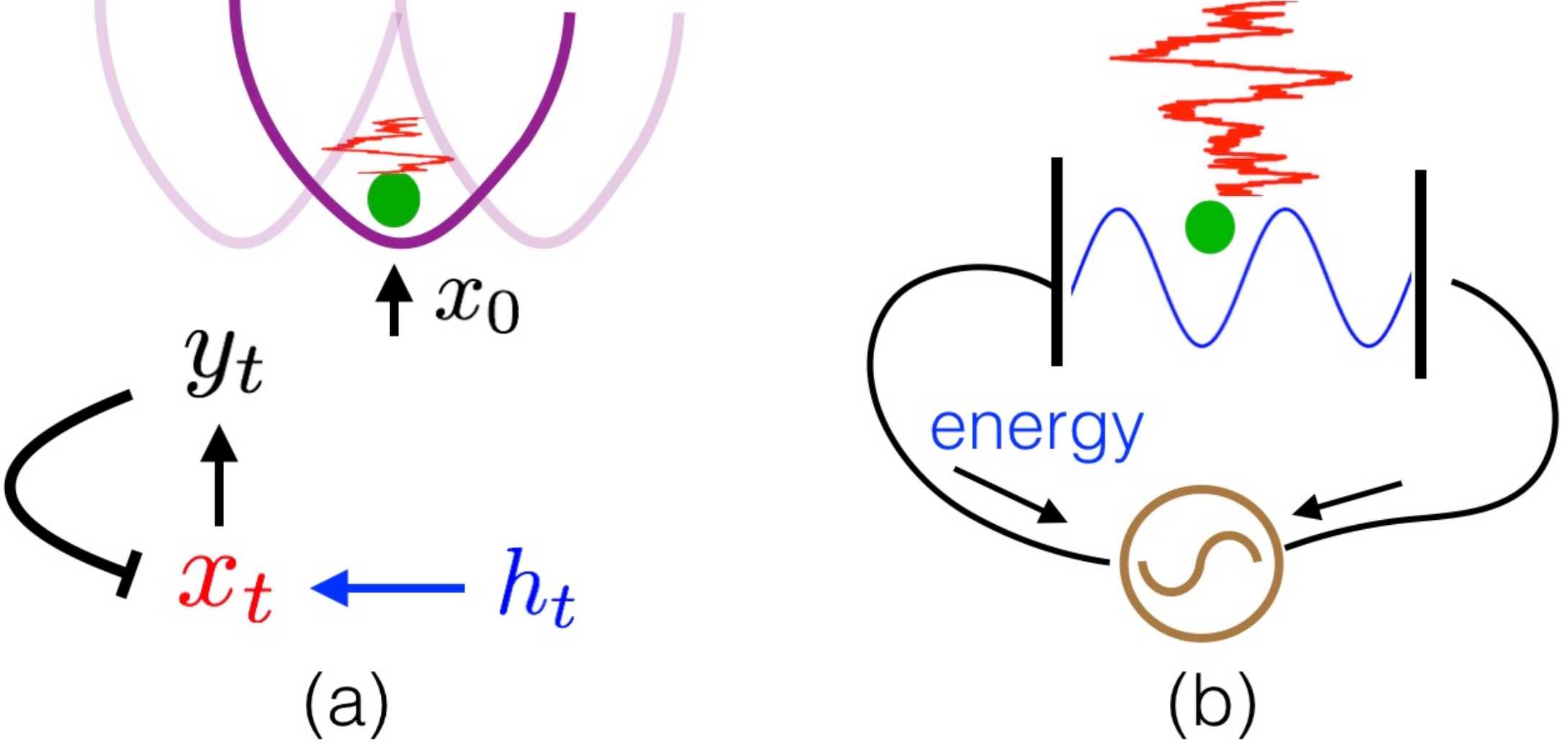}
\caption{(a) Generating negative response.   A green particle is trapped within a potential.  The location of the potential minimum, $y_t$,  is varied through an integral feedback control to ensure that the particle stays around $x_0$, regardless of a slow external perturbation $h_t$.    (b)  The frequency-resonant engine.  Assuming that the particle is charged,  we can extract its energy  by applying  through the alternating current (AC) source an  electric  field  $h_t$,  stochastic or not, with a correlation spectrum   localized where the particle has  negative response.  } 
\label{fig:schemes}
\end{figure}

 A main strategy to achieve adaptation, thus negative response, is to implement  integral feedback control through  an internal variable $y$~\cite{yi2000robust},   as  illustrated  in  FIG.~\ref{fig:schemes}(a).   In some sense,   Maxwell's demon is still required for extracting energy.  However,  it  no longer controls the external operation,  but rather the internal response.   Given  a charged particle that is trapped in such a way,   we may further apply an  alternating electric field with an optimized frequency spectrum to extract its energy,  as illustrated in  FIG.~\ref{fig:schemes}(b).  See~\cite{toyabe2010experimental} for a similar experimental setup.

These observations  are valid for more general blind external operations. Let us consider 
\begin{equation}
\gamma \dot{x}=F(x,{\vec y},t)-\partial_xU(x,h)+\xi,
\label{eq:main-langevin-general}
\end{equation} 
with $U(x,h)$ the coupling energy between $x$ and $h$.     Here,  we   introduce the conjugate variable $B_t\equiv -\partial_h U(x_t,h^*)$,  which is assumed to be bounded.    The average energy injection rate  through the external operation is given by $\dot{W}_{ext}^*=\langle  (-B_t)\circ \dot{h}_t \rangle_*$, or  $\langle  \dot{B}_t \circ h_t \rangle_*$ due to  integration by parts.    Introducing the linear response function $R_{\dot{B}}(t-\tau)\equiv \delta \langle \dot{B}_t\rangle/\delta h_\tau$,  we obtain
\begin{equation}
\dot{W}_{ext}^*=  \int_{-\infty}^\infty \tilde{R}_{\dot{B}}'(\omega)\tilde{C}^*_h(\omega)\frac{d\omega}{2\pi},
\label{eq:Wext-general}
\end{equation}
 valid up to the second order of $h$.     When $U=-xh$,  we have $\dot{B}_t=\dot{x}_t$,  which reproduces our previous result.  For  $U=hx^2/2$,  which is also commonly implemented   experimentally~\cite{Martinez2016Brownian,Rosnagel2016atomengine},   we have $\dot{B}_t=-\partial_t(x_t^2/2)$.

\emph{Potential shifting model.}--Below,  we demonstrate our main results through the following  model :
\begin{subequations}\label{eq:adaptation}
\begin{eqnarray}
\gamma \dot{x}&=&-\partial_x U_0(x-b y)+h+\xi\\
\gamma_1 \dot{y} &=&- k_1(x-x_0)-k_2 y+\xi_1. 
\end{eqnarray}
\end{subequations}
Here,  $\gamma_1$ is the friction coefficient for $y$.  Both $\xi$ and $\xi_1$ are zero-mean Gaussian white noises, satisfying $\langle \xi(t)\xi(\tau)\rangle=2\gamma T \delta(t-\tau)$ and $\langle \xi_1(t)\xi_1(\tau)\rangle=2\gamma_1 T\delta(t-\tau)$.     $U_0$ should be a single-well potential.   This model   may describe a  colloid particle trapped by a laser,  whose position $y_t$  is controlled dynamically with respect to that of the particle,  as illustrated in FIG.~\ref{fig:schemes}(a).   For $U_0(x-by)=k(x-by)^2/2$,  we show in the Supplemental Material~\cite{supp}  that the response spectrum $\tilde{R}_{\dot{x}}'(\omega)$ becomes negative in the frequency region  $\omega<\sqrt{\omega_x\omega_y-\omega_2^2}$,
where $\omega_x\equiv k/\gamma $,  $\omega_y\equiv b k_1/\gamma_1$, and $\omega_2\equiv k_2/\gamma_1$.  This is possible only when $\omega_2<\sqrt{\omega_x\omega_y}$,  i.e., the restoring force $-k_2y$ should be  relatively small.  
  Below,  we focus on a harmonic potential $U_0$, and take $k_2=0$  for simplicity.   This system is characterized by a dimensionless quantity  $\omega_y/\omega_x$,  which measures the relative feedback speed of the system.

\begin{figure}
\centering
\includegraphics[width=8.5cm]{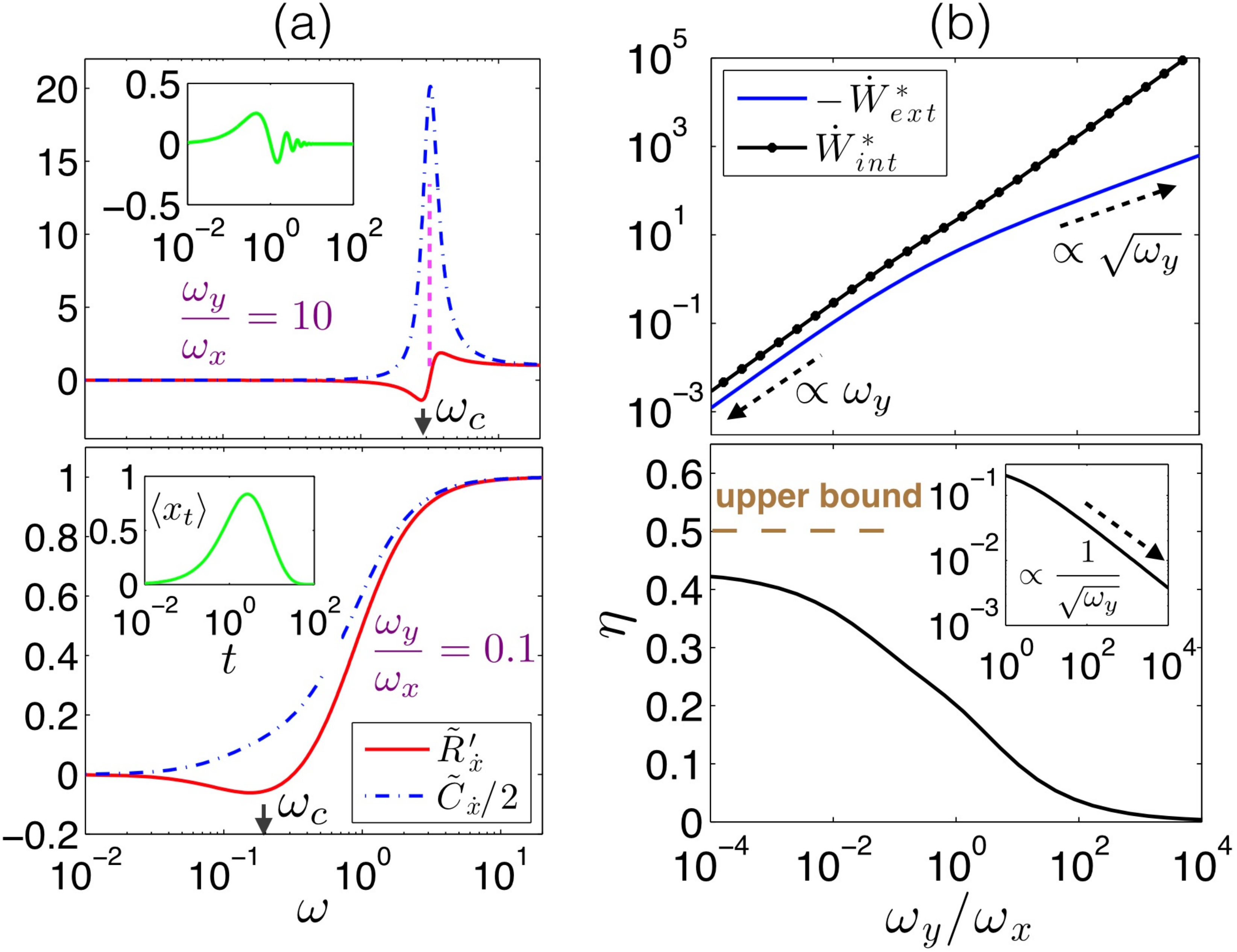}
\caption{ (a)  The velocity response spectrum $\tilde{R}_{\dot{x}}'(\omega)$ and correlation spectrum $\tilde{C}_{\dot{x}}(\omega)$ of the linearized  potential shifting model.  The green curves in the insets give the response $\langle x_t\rangle$ to the perturbation $h\to h+1$.   (b) The externally extracted power $-\dot{W}^*_{ext}$,  the internally supplied power $\dot{W}^*_{int}$,  and the efficiency $\eta$ at various relative feedback speed $\omega_y/\omega_x$,  when the system is perturbed with $h_t=5\cos(\omega_c t)$.        Other parameters: $\omega_x=1$,  $\gamma=1$,  $T=1$,  $k_1=1$, $b=1$ and $k_2=0$. }
\label{fig:response}
\end{figure}

Let us first study the temporal response  of this system to external perturbations.    For a step jump $h\to h+1$  at $t=0$,  the system evolves as   
\begin{equation}\label{eq:xt}
\langle x_t\rangle=x_0+\frac{1}{\gamma} \frac{\exp(-\lambda_- t)-\exp(-\lambda_+t)}{ \lambda_+-\lambda_-},
\end{equation}
 with $\lambda_\pm\equiv \frac{1}{2}\omega_x (1\pm \sqrt{1-4 \omega_y/\omega_x})$.   Since the real part of  $\lambda_\pm$ is always  positive,  the elicited  deviation from $x_0$  will eventually vanish. Therefore, this model with $k_2=0$ has  achieved  perfect adaptation.   In the long run,  the variation of $h$ needs to be  compensated by that of $y$.

 According to Eq.~(\ref{eq:xt}),  returning  to $x_0$ could be either oscillatory,  when a relatively fast feedback is implemented, i.e., $\omega_y/\omega_x\gg 1$,  or  steady for a slow feedback.  These are shown by the green curves in the insets of FIG.~\ref{fig:response}(a).     Both strategies have been exploited by nature for different purposes.     The hair bundle works in the fast feedback region with an oscillatory response in order to amplify an auditory signal at a specific frequency~\cite{Hudspeth2008hair}.  On the other hand,   the sensory circuit in E. \emph{coli} works in the slow feedback region without oscillation,  which is suitable for detecting all fast variation of the extracellular ligand concentration~\cite{tu2013quantitative}.

Solving  Eq.~(\ref{eq:adaptation}) in the Fourier space,  we obtain the desired velocity response spectrum:
\begin{equation}
\tilde{R}'_{\dot{x}}(\omega)=\frac{1}{\gamma} \frac{\omega^2 (\omega^2-\omega_x\omega_y)}{(\omega^2-\omega_x\omega_y)^2+\omega^2\omega_x^2}.
\label{eq:Rx-model}
\end{equation}
It reaches zero at $\omega\to 0$. Therefore,   a sufficiently slow perturbation is energetically reversible.  On the other hand,  it approaches $1/\gamma$ for $\omega\to \infty$,  which implies that a super fast perturbation, to which  the system is almost insensitive,  is highly dissipative.   This marks a sharp distinction between energetics and dynamics of a non-equilibrium system,  which is quite general~\cite{hondou2000unattainability, celani2012anomalous,esposito2012stochastic, kawaguchi2013fluctuation,bo2014entropy,jia2016model,Wang2016entropy,Wang2016FRRviolation}.   

  The spectrum becomes negative in the low frequency region $\omega<\sqrt{\omega_x\omega_y}$, as shown through the red curves in    FIG.~\ref{fig:response}(a), which allows energy extraction.  In the lower panel of  FIG.~\ref{fig:response}(a),  obtained at $\omega_y/\omega_x=0.1$,    the negative spectrum is quite small over a broad  frequency region.   For the fast feedback case,   the response spectrum becomes  much  larger within a narrow frequency window,  as shown in the upper panel of FIG.~\ref{fig:response}(a).  The minimum value is reached at   $\omega_c=\sqrt{\omega_x\omega_y/(1+\sqrt{\omega_x/\omega_y})}$,  which is optimal for extracting energy using a periodic operation.   The velocity correlation spectrum $\tilde{C}_{\dot{x}}$ is also illustrated in FIG.~\ref{fig:response}(a).  The FDT  is violated significantly around $\omega=\omega_c$,  and  restored approximately in the high frequency region.  At  $\omega_y/\omega_x=10$,  the correlation spectrum exhibits a sharp peak at the frequency that is slightly larger than $\omega_c$.  Counterintuitively,  perturbation at such a characteristic frequency is dissipative, as indicated by the vertical dashed line in the upper panel of  FIG.~\ref{fig:response}(a).

Below,  we analyze the energetics for this linear model under any perturbation.  The averaged dissipation rate through the frictional motion of  $x$ is given by $\dot{Q}^*_x\equiv \langle [\gamma \dot{x}_t-\xi(t)]\circ \dot{x}_t \rangle_*$~\cite{sekimoto2010stochastic}.  Inserting  $\gamma \dot{x}_t-\xi(t)=-k(x_t-b y_t)+h_t$ and noting that $\langle x_t\circ \dot{x}_t\rangle_*=0$,   we obtain $\dot{Q}^*_x=b k\langle y_t\circ \dot{x}_t\rangle_*+\dot{W}^*_{ext}$.  Similarly,  the averaged  dissipation rate for $y$ satisfies $\dot{Q}^*_y\equiv  \langle [ \gamma_1 \dot{y}_t-\xi_1(t)]\circ \dot{y}_t \rangle_*=-k_1\langle x_t\circ \dot{y}_t\rangle_*$.   From integration by parts,  we obtain $-\langle x_t\circ \dot{y}_t\rangle_*=\langle y_t\circ \dot{x}_t\rangle_*$.   Noting the energy conservation  $\dot{Q}^*_x+\dot{Q}^*_y=\dot{W}^*_{ext}+\dot{W}^*_{int}$,  with $\dot{W}^*_{int}$ being the power supplied by an internal energy source,   we finally obtain the following relations
\begin{equation}
\dot{Q}^*_x=\frac{bk}{k_1} \dot{Q}^*_y+\dot{W}^*_{ext}, \quad \dot{W}^*_{int} =\frac{bk+k_1}{k_1} \dot{Q}^*_y,
\label{eq:Qx-Wext}
\end{equation}
valid for any $h_t$.  Therefore,  the efficiency is given by 
\begin{equation}
\eta\equiv \frac{-\dot{W}^*_{ext}}{\dot{W}^*_{int}}=\frac{bk}{bk+k_1}-\frac{k_1}{bk+k_1}\frac{\dot{Q}^*_x}{\dot{Q}^*_y}\le \frac{1}{1+(k_1/bk)},
\label{eq:upperbound}
\end{equation}
where we have used the constraint that dissipation is always positive, i.e.,  $\dot{Q}^*_x\ge 0$ and $\dot{Q}^*_y\ge 0$,  since $h_t$ contains no information about this system.   Eq.~(\ref{eq:upperbound}) gives the upper bound of the efficiency,  determined only by $k_1/bk$,  the coupling asymmetry between $x$ and $y$,  which remains valid even when $k_2\neq 0$.

Now,  we consider perturbation at the optimal frequency $\omega_c$, i.e., $h_t=h_0\cos (\omega_c t)$.  In this case, $\tilde{C}_h^*(\omega)=\frac{\pi}{2}h_0^2 [\delta (\omega+\omega_c)+\delta(\omega-\omega_c)]$.  Combined with Eq.~(\ref{eq:Rx-model}) and Eq.~(\ref{eq:spectrum}) and noting that  $\alpha=0$,        we obtain \begin{equation}
-\dot{W}^*_{ext}=\frac{h_0^2}{2\gamma} \frac{\omega_y/\omega_x}{1+2\sqrt{\omega_y/\omega_x}}.
\end{equation}
In the Supplemental Material~\cite{supp},  we calculate the rate of internal energy supply:
\begin{equation}
\dot{W}^*_{int}=\frac{h_0^2\omega_y}{2\gamma\omega_x}(1+\frac{k_1}{bk})\theta+ T\omega_y(2+\frac{bk}{k_1}+\frac{k_1}{bk}),
\label{eq:main-W-int}
\end{equation}
where $\theta=(\sqrt{\omega_y/\omega_x}+1)/(2\sqrt{\omega_y/\omega_x}+1)$,  ranging between $[0.5, 1]$.  At $h_0=0$,  $\dot{W}^*_{int}\ge 4\omega_yT$, with the equality achieved at symmetric coupling: $bk=k_1$.  This irreducible energetic cost is needed to maintain  negative response.     Although  a larger  perturbation amplitude  increases  the output power,  it also requires a larger internal energy supply.  Interestingly,   both the output power and efficiency can be increased simultaneously by a larger $h_0$.

    In FIG.~\ref{fig:response}(b),   we illustrate how the extracted power and the internal energy supply depend on the feedback speed $\omega_y$,  rescaled by $1/\omega_x$.   The extracted power increases linearly with $\omega_y$ when the feedback is relatively slow,  but scales with $\sqrt{\omega_y}$ when the feedback becomes faster.   On the other hand,  the  internal energy supply  always grows with $\omega_y$. Therefore,  the efficiency $\eta\propto 1/\sqrt{\omega_y}$  in the fast feedback region,  as shown in the inset in FIG.~\ref{fig:response}(b).  In the slow feedback region,  the efficiency is maximized,  bounded from above  by $bk/(bk+k_1)$,  as shown by the brown dashed line in  FIG.~\ref{fig:response}(b).  This upper bound could be achieved in the limit  $h_0\to \infty$ and $\omega_y/\omega_x\to 0$ .  In general,    the output power and efficiency cannot be optimized simultaneously,   as in the case of heat engines.

   \emph{ Concluding Remarks.}--Given a system that is powered  out of equilibrium by an internal energy source,  it seems possible to extract its energy that would otherwise be dissipated.  Here,   considering a blindly applied external operation,    we find that  negative response in a certain frequency region  is necessary for energy extraction,  and the spectrum of the external operation must be localized at this  region.    Negative response  could be achieved with an internal variable  that implements  integral feedback control.   Remarkably,  energy extraction has already been demonstrated with hair bundles in the inner ear~\cite{Martin1999hairResponse}.  With the physics  clarified here,  we expect  simpler experimental demonstration   in the  near future.

   We also find that   adaptation to a slow external operation necessitates negative response in a certain frequency region. Therefore,  adaptation of sensory systems must be powered by an internal energy supply even without signal variation,  as  has been conjectured   previously in a case study for the sensory circuit in E. \emph{coli}~\cite{lan2012energy,shouwen2015adaptation}.    Our  argument is  based on causality,  and the  assumption that  the external signal only affects the output directly,  as valid for both the sensory circuit in E. \emph{coli} and the hair bundles.  It remains to be explored whether other response behavior will  also cost energy.

 \begin{acknowledgements}

The authors thank Kyogo Kawaguchi,   Ganhui Lan and Stefano Bo  for their  helpful discussion on this work.    
The work was supported in part by the NSFC under Grant No. U1430237 and by the Research Grants Council of the Hong Kong Special Administrative Region (HKSAR) under Grant No. 12301514. 
\end{acknowledgements}



\def\theequation{S\arabic{equation}}
\makeatletter
\@addtoreset{equation}{section}
\makeatother

\setcounter{equation}{0}

\def\thefigure{S\arabic{figure}}
\makeatletter
\@addtoreset{figure}{section}
\makeatother
\setcounter{figure}{0}
\newcommand{\wh}[1]{\widehat{#1}}
\newpage

\onecolumngrid

\section{Supplemental Material}

\twocolumngrid

\subsection{Derivation of Eq.~(\ref{eq:spectrum}) in the Main Text}

To apply an external force $f_t$ given by Eq.~(\ref{eq:ft}) in the Main Text,  we need to dissipate energy at the rate
\begin{equation}
\dot{W}_{ext}^* =(1-\alpha) \dot{W}_h^*+\alpha  \dot{W}_{fb}^*,
\label{eq:temp-W}
\end{equation}
where $\dot{W}_h^*\equiv \langle h_t\circ \dot{x}_t\rangle_*$ is the contribution from blind perturbation while $\dot{W}_{fb}^*\equiv -\langle \int_{-\infty}^{\infty} G(t-\tau) \dot{x}_\tau d\tau \circ \dot{x}_t \rangle_*$ is from the information-based operation.  

 Replacing $\tau$ by $s+t$,  we have
 \begin{eqnarray}
\dot{W}_{fb}^*& =&- \int_{-\infty}^{\infty} G(-s) \langle \dot{x}_{s+t}  \circ \dot{x}_t \rangle_* ds\nonumber\\ 
&=&-\int_{-\infty}^\infty\int_{-\infty}^\infty G(-s) [C^*_{\dot{x}}(s)+2\pi v_0 \delta(s) ]ds \nonumber\\
&=&-\tilde{G}'(0)v_0- \int_{-\infty}^\infty \tilde{G}'(\omega)\tilde{C}_{\dot{x}}^*(\omega)\frac{d\omega}{2\pi}.\label{eq:feedback-W}
\end{eqnarray}
The last equality is obtained by replacing $G(-s)$ and $C^*_{\dot{x}}(s)$ by their corresponding Fourier transform, i.e.,  $G(-s)=\int_{-\infty}^\infty \tilde{G}(\omega) \exp(-i\omega [-s]) d\omega/2\pi$ and $C^*_{\dot{x}}(s)=\int_{-\infty}^\infty \tilde{C}_{\dot{x}}^*(\omega_1) \exp(-i\omega_1 s) d\omega_1/2\pi$,  and integrating over $s$ and $\omega_1$.   The imaginary part of $\tilde{G}(\omega)$,  denoted as  $\tilde{G}''(\omega)$,  does not appear since it is an odd function while 
$\tilde{C}_{\dot{x}}^*(\omega)$ is an even function.   As mentioned in the Main Text,    $\tilde{C}_{\dot{x}}^*$ depends on how the system is perturbed,  which is different from $\tilde{C}_{\dot{x}}$, obtained with $h=0$.  

On the other hand,  treating $\delta h_t=h_t-h^*$ as a small perturbation to the system,  the leading order response is given by  
\begin{equation}
\langle \dot{x}_t\rangle=\langle \dot{x}\rangle_{ss}+\int_{-\infty}^\infty R_{\dot{x}}(t-\tau) \delta h_\tau d\tau.
\end{equation}
Here,   $\langle \dot{x}\rangle_{ss}$ is the steady-state mean velocity before applying perturbation,  which may be different from $v_0$,  and $A$ characterizes the amplitude of the perturbative force $h$.      Therefore, 
\begin{eqnarray}
\dot{W}_h^* &=& \langle \dot{x}\rangle_{ss} h^*+\Big\langle \delta h_t  \int_{-\infty}^\infty R_{\dot{x}}(t-\tau) \delta h_\tau d\tau \Big\rangle_* \nonumber\\
 &=&\langle \dot{x}\rangle_{ss} h^*+ \int_{-\infty}^\infty \tilde{R}_{\dot{x}}'(\omega)\tilde{C}_h^*(\omega)\frac{d\omega}{2\pi}, \label{eq:perturbation-W}
\end{eqnarray}
which gives the first and second order contribution explicitly.  Inserting Eq.~(\ref{eq:feedback-W}) and Eq.~(\ref{eq:perturbation-W}) back into Eq.~(\ref{eq:temp-W}),  we finally obtain 
\begin{equation}
\begin{split}
&\dot{W}_{ext}^*=\alpha h^* \langle \dot{x}\rangle_{ss} -(1-\alpha) \tilde{G}'(0) v_0+\\
&\int_{-\infty}^\infty \Big (\alpha  \tilde{R}_{\dot{x}}'(\omega)\tilde{C}_h^*(\omega)- (1-\alpha) \tilde{G}'(\omega)\tilde{C}_{\dot{x}}^*(\omega)\Big)\frac{d\omega}{2\pi}.\\
\end{split}
\end{equation}
In the case without   drifting motion,  i.e., $\langle \dot{x}\rangle_{ss} =\langle \dot{x}\rangle_*=0$,  we obtain Eq.~(\ref{eq:spectrum}) in the Main Text.

Below,  we discuss the property of the correlation spectrum.  Firstly, 
\begin{equation}
\begin{split}
\tilde{C}_h^*(\omega) =&\int_{-\infty}^\infty C_h^*(s)\exp(i\omega s) ds\nonumber\\
=& \lim_{\mathcal{T}\to\infty} \frac{1}{2\mathcal{T}} \int_{-\mathcal{T}}^\mathcal{T} \int_{-\mathcal{T}}^\mathcal{T} \delta h_t \delta h_{t+s} \exp(i\omega s) dt ds,\nonumber\\
\end{split}
\end{equation}
which gives 
\begin{equation}
\tilde{C}_h^*(\omega)=\tilde{\delta h}^+(\omega)\tilde{\delta h}^+(-\omega)=  |\tilde{\delta h}^+(\omega)|^2\ge 0. 
\end{equation}
At $\omega\neq 0$,  it becomes $\tilde{C}_h^*(\omega)=  |\tilde{ h}^+(\omega)|^2\ge 0$ and at $\omega=0$,   $\tilde{C}_h^*(0)=0$.   

Similarly,  for $\omega\neq 0$,  the velocity correlation spectrum satisfies 
\begin{equation}
\tilde{C}_{\dot{x}}^*(\omega)=\langle \tilde{\dot{x}}^+(\omega)\tilde{\dot{x}}^+(-\omega)\rangle=\omega^2 \langle |\tilde{x}(\omega)|^2\rangle\ge 0,
\end{equation}
where ensemble average is needed to deal with the stochastic nature of the trajectory $\dot{x}_t$.  To arrive at the  second equality,   we have used $\tilde{\dot{x}}(\omega)=-i\omega\tilde{x}(\omega)$.   At $\omega=0$,  we have
\begin{eqnarray}
\tilde{C}_{\dot{x}}^*(0)&=&\lim_{\mathcal{T}\to \infty}  \int_{-\mathcal{T}}^{\mathcal{T}} \langle (\dot{x}_{t+s}-v_0)(\dot{x}_t-v_0)\rangle_* ds \nonumber  \\
&=&2 \bar{D}\ge 0, 
\label{eq:green-kubo}
\end{eqnarray}
with $\bar{D}$ being the effective diffusion coefficient in the large time limit.  This is due to  the Green-Kubo relation~\cite{kubo1966fluctuation}.   If there is no drifting velocity, i.e., $v_0=0$,  we have $\bar{D}=0$.

\subsection{Generating negative response in a 2-d system}
Consider a linear version of Eq.~(\ref{eq:adaptation}) in the Main Text:
\begin{equation}
\gamma \dot{x}=-kx+bky+h+\xi,\quad \gamma_1\dot{y}=-k_1 x-k_2 y+\xi_1.
\label{eq:linearModel-0}
\end{equation}
Fourier transforming these equations give
\[ -i\omega \gamma \tilde{x}=-k\tilde{x}+bk\tilde{y}+\tilde{h}+\tilde{\xi},\;\;  -i\omega \gamma_1 \tilde{y}=-k_1\tilde{x}-k_2\tilde{y}+\tilde{\xi}_1. \]
With $\omega_x\equiv k/\gamma$, $\omega_y\equiv bk_1/\gamma_1$ and    $\omega_2\equiv k_2/\gamma_1$, we have  
\begin{equation}
\tilde{x}=\frac{1}{\gamma} \frac{ (i\omega-\omega_2)(\tilde{h}+\tilde{\xi})-  \tilde{\xi}_1 bk/\gamma_1}{ \omega^2-\omega_2\omega_x-\omega_x\omega_y +i\omega(\omega_x+\omega_2) },
\label{eq:x-fre-k2}
\end{equation}
  The velocity response spectrum is then given by 
\begin{eqnarray}
\tilde{R}_{\dot{x}}(\omega)&=&-i\omega \Big \langle\frac{\delta \tilde{x}(\omega)}{\delta \tilde{h}(\omega)} \Big\rangle\nonumber\\
&=&  \frac{1}{\gamma} \frac{ \omega^2+i\omega \omega_2}{  \omega^2-\omega_2\omega_x-\omega_x\omega_y +i\omega(\omega_x+\omega_2)},
\end{eqnarray}
and its real part is 
\begin{equation}
\tilde{R}_{\dot{x}}'(\omega)= \frac{1}{\gamma} \frac{ \omega^2(\omega^2-\omega_x\omega_y+\omega_2^2) }{(\omega^2- \omega_2\omega_x -\omega_x \omega_y)^2 +\omega^2(\omega_x+\omega_2)^2}.
\end{equation}
Therefore, the response becomes negative in the  frequency region
\begin{equation}
\omega<\sqrt{\omega_x\omega_y-\omega_2^2},
\label{eq:requirement-omega3}
\end{equation}
which requires that $\omega_2<\sqrt{\omega_x\omega_y}$, i.e.,  a relatively small restoring force.

\subsection{Derivation of  Eq.~(\ref{eq:main-W-int}) in the Main Text}
Here,  we consider the model introduced in Eq.~(\ref{eq:linearModel-0}) at $k_2=0$.   Similarly as Eq.~(\ref{eq:x-fre-k2}),  we obtain 
\begin{equation}
\tilde{x}^+(\omega)=\frac{1}{\gamma}\frac{i\omega (\tilde{h}^++\tilde{\xi}^+)- \tilde{\xi}_1^+ bk/\gamma_1}{\omega^2 -\omega_x\omega_y+i\omega \omega_x},
\label{eq:rescaled-x}
\end{equation}
where $\tilde{x}^+(\omega)$ is the rescaled Fourier transform of $x_t$.   Its  velocity correlation spectrum at $h=0$ is given by 
\begin{eqnarray}
\tilde{C}_{\dot{x}}(\omega)&=&\omega^2 \langle |\tilde{x}^+(\omega)|^2\rangle\nonumber \\
&=&\frac{2}{\gamma}\frac{  \omega^2(\omega^2+\omega_x b^2 k/\gamma_1) }{(\omega^2 -\omega_x\omega_y)^2+\omega^2 \omega_x^2}. 
\end{eqnarray}
where we have used   $\langle |\tilde{\xi}^+(\omega)|^2\rangle=2T\gamma$ and $\langle |\tilde{\xi}_1^+(\omega)|^2\rangle=2T\gamma_1$. 

Now, we proceed to  calculate  $\dot{W}_{int}^*$, the rate of  internal energy supply.      Firstly,  the dissipation rate through frictional motion of $y_t$ is given by 
 \begin{equation}
 \dot{Q}^*_y=-k_1\langle x_t\circ \dot{y}_t\rangle_*= \frac{k_1^2}{\gamma_1}\langle x_t^2\rangle_*.
 \end{equation}
   In the second equality we have replaced $\dot{y}$ by $(-k_1x+\xi_1)/\gamma_1$ and used $\langle x_t\circ \xi_1(t)\rangle_*=0$.  Combined with Eq.~(\ref{eq:Qx-Wext}) in the Main Text,  we obtain  the rate of internal energy supply:  
\begin{equation}
\dot{W}_{int}^*=\omega_y (k+k_1/b) \langle x_t^2\rangle_*.
\label{eq:diss-W-int}
\end{equation}
The long-time averaged variance $\langle x_t^2\rangle_*$ is given by 
\begin{equation}\label{eq:x-var}
\langle x_t^2\rangle_* =\int_{-\infty}^\infty \langle \tilde{x}^+(\omega)\tilde{x}^+(-\omega)\rangle\frac{d\omega}{2\pi},
\end{equation}
Inserting $\tilde{x}^+(\omega)$ from  Eq.~(\ref{eq:rescaled-x}),  we have  
\begin{eqnarray}
\langle x_t^2\rangle_*&=&\int_{-\infty}^\infty \frac{1}{\gamma^2} \frac{\omega^2 [\tilde{C}_h^*(\omega)+2T\gamma]+2T\gamma_1 (bk/\gamma_1)^2}{(\omega^2-\omega_x\omega_y)^2+\omega^2\omega_x^2}\frac{d\omega}{2\pi}\nonumber\\
&=& \int_{-\infty}^\infty \frac{1}{\gamma^2} \frac{\omega^2 \tilde{C}_h^*(\omega)}{(\omega^2-\omega_x\omega_y)^2+\omega^2\omega_x^2}\frac{d\omega}{2\pi}+\frac{T}{k}+\frac{bT}{k_1},\nonumber\\
\label{eq:fluctuation}
\end{eqnarray}
where we have used the following integrals
\begin{equation}
\int_{-\infty}^\infty  \frac{\omega^2}{(\omega^2-\omega_x\omega_y)^2+\omega^2\omega_x^2}\frac{d\omega}{2\pi}=\frac{1}{2\omega_x},
\end{equation}
\begin{equation}
\int_{-\infty}^\infty  \frac{1}{(\omega^2-\omega_x\omega_y)^2+\omega^2\omega_x^2}\frac{d\omega}{2\pi}=\frac{1}{2\omega_x^2\omega_y}.
\end{equation}
To proceed,  we need to specify the perturbation spectrum. Let us consider a periodic perturbation $h_t=h_0\cos (\omega_h t)$. Its   correlation spectrum is 
\begin{equation}
\tilde{C}_h^*(\omega)= \frac{\pi h_0^2}{2}\left [\delta (\omega+\omega_h)+\delta (\omega-\omega_h)\right].
\end{equation}
 Plugging it into Eq.~(\ref{eq:fluctuation}),  we obtain
\begin{equation}
\langle x_t^2\rangle_*=  \frac{h_0^2}{2\gamma^2} \frac{\omega_h^2 }{(\omega_h^2-\omega_x\omega_y)^2+\omega_h^2\omega_x^2}+ \frac{T}{k}+\frac{bT}{k_1}.
\end{equation}
Therefore,  according to Eq.~(\ref{eq:diss-W-int}),  we have 
\begin{equation}
\begin{split}
\dot{W}^*_{int}&=\omega_y\omega_x (1+\frac{k_1}{bk})\frac{ h_0^2}{2\gamma}   \frac{\omega_h^2 }{(\omega_h^2-\omega_x\omega_y)^2+\omega_h^2\omega_x^2}\\
&\quad +T \omega_y(2+\frac{bk}{k_1}+\frac{k_1}{bk}).
\end{split}
\label{eq:W-int-2}
\end{equation}
At the optimal frequency $\omega_c=\sqrt{\omega_x\omega_y/[1+\sqrt{\omega_x/\omega_y}]}$,  we finally obtain Eq.~(\ref{eq:main-W-int}) in the Main Text.

\end{document}